PUBLISHED IN DATA IN BRIEF (ELSEVIER): https://doi.org/10.1016/j.dib.2025.112175

# Mapping the Invisible Internet: Framework and Dataset

**Siddique Abubakr Muntaka[a,*], Jacques Bou Abdo [a,*], Kemi Akanbi [a], Sunkanmi Oluwadare [a], Faiza Hussein [b], Oliver Kornyo [c], Michael Asante [c]**

[a]*University of Cincinnati, School of Information Technology, Cincinnati, Ohio, USA*
[b]*Garden City University College, Computer Science Department, Kumasi, Ashanti, Ghana*
[c]*Kwame Nkrumah University of Science and Technology, Department of Computer Science, Ghana*
muntaksr@mail.uc.edu, bouabdjs@ucmail.uc.edu

**Abstract**

This article describes a novel dataset that maps the network layer of the Invisible Internet Project (I2P). The data was collected using SWARM-I2P framework, which deployed I2P routers as a network of mapping agents that gather information on the network's topology and traffic over an extended period. The dataset documents over 50,000 nodes, including subsets of high-performance (FastSet) nodes and high-capacity nodes characterized by metrics such as bandwidth, latency, and uptime. It also contains detailed records of network traffic and the geographic distribution of thousands of nodes. Data was collected using a combination of methods, including querying router consoles, analysing the network database (netDb), and passive network monitoring. All node identifiers were anonymized to maintain user privacy. The data is publicly available in CSV and TXT formats on Zenodo, with mapping scripts provided on GitHub. This resource provides a foundational understanding of the decentralized routing behaviours that underpin I2P's anonymity, making it suitable for reuse in analyses of tunnel node selection, anonymity network resilience, and adversarial modelling.

## SPECIFICATIONS TABLE

| | |
|---|---|
| **Subject** | Computer Sciences |
| **Specific subject area** | Evaluation of I2P router node selection for building and maintaining tunnels. |
| **Type of data** | Text files in CSV, txt format and Excel. |
| **Data collection** | We collected over 50,000 node identifiers from I2P's network using standard and floodfill routers. Standard routers captured node connections and tunnel formation, while the floodfill routers captured broader network patterns. We recorded activity frequency, network role, location, and capacity, enabling a detailed study of I2P's routing behaviours. |
| **Data source location** | Multiple global locations of routers on the Garlic network (I2P Network). |
| **Data accessibility** | Repository name: Zenodo<br>Data identification number: 10.5281/zenodo.15278912<br>Direct URL to data: https://zenodo.org/records/15369068<br>Instructions for accessing these data: Publicly available via the link. |
| **Related research article** | None |

# VALUE OF THE DATA

- This dataset provides a large-scale mapping of the I2P network layer, capturing over 50,000 nodes used in tunnel formation, thus focusing on the network layer. This provides novel insights into the node selection dynamics of the garlic routing system.
- It enables detailed analysis of decentralized routing mechanisms by documenting dynamic interactions among nodes, thereby supporting studies on tunnel efficiency and the robustness of anonymity networks. Unlike prior application-layer datasets that primarily focus on discovering hidden services like eepsites, our dataset's unique focus on the network layer enables researchers to address fundamental questions about the emerging structure that helps understand the attacker ratio, network resilience, anonymity, and node selection that are unobservable in application-layer data.
- The data includes temporal and spatial markers, allowing researchers to examine the influences of regional clustering and network topology on node selection, which is critical for understanding communication resilience.
- It provides a resource for validating network models mathematically and computationally, as well as simulating attack surfaces, thereby supporting cybersecurity research and the optimization of privacy-preserving protocols.
- Owing to the comprehensiveness and fine-grained structure of this dataset, it constitutes a valuable resource for network scientists, law-enforcement agencies (LEAs), cybersecurity researchers, and developers seeking to analyze and enhance decentralized systems and anonymous communication networks (ACNs), such as the Invisible Internet Project (I2P).

# BACKGROUND

The Invisible Internet Project (I2P) is an anonymous overlay network [1], similar to the Onion Router (Tor), that facilitates private communication through layered encryption and decentralized routing [1,2]. Unlike traditional web protocols, I2P employs garlic routing. This mechanism encapsulates multiple messages (cloves) within a single encrypted message, known as a garlic bulb [3], which enhances traffic obfuscation and reduces correlation risks between the sender and recipient. Existing studies predominantly focus on the application layer of I2P, such as discovering hidden services like eepsites (I2P site) [4]. At the same time, the underlying peer-to-peer dynamics responsible for tunnel construction remain largely underexplored. For this study, the terms "node" and "peer" are used interchangeably to refer to a participant in the I2P network. The mechanism for constructing tunnels through node selection is fundamental to the operation of the I2P; hence, an open research question is “Is there a way that I2P could perform peer selection more efficiently or securely? [5]”. Understanding current operations and selecting nodes is essential, as it involves analyzing the nodes specified in tunnels. In an I2P session, each participant creates an inbound and an outbound tunnel. Consequently, when two nodes communicate, four unidirectional tunnels are formed. As illustrated in Fig 1, if Alice as client user wishes to send/receive data from Bob, she obtains Bob’s inbound tunnel gateway from the network database (netDb). Alice then forwards her encrypted message through her outbound tunnel to Bob’s inbound gateway. After decryption, Bob’s gateway delivers the message to his router (I2P router). Bob responds by mirroring this process in reverse. The volunteer nodes in these tunnels are selected based on their profile, considering bandwidth capacity, reliability, uptime, and latency. This process is known as the reputation mechanism [6].

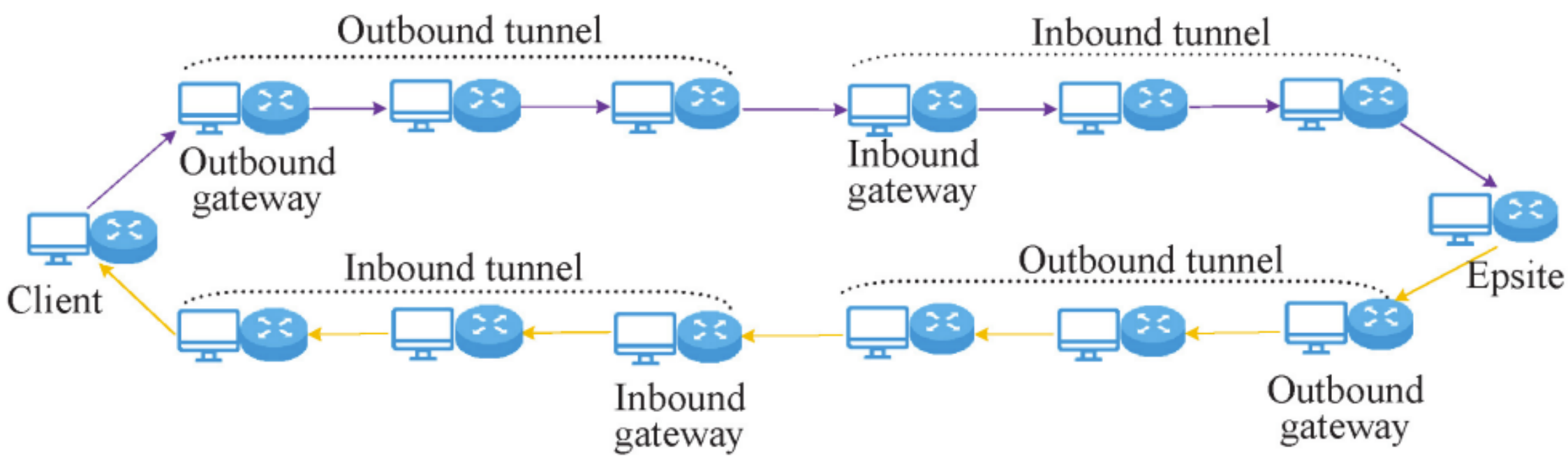

Fig 1. Communication between two I2P Nodes [7]

Mapping this dynamic topology poses considerable challenges. The network continuously change due to the spontaneous joining and leaving of nodes [8]. According to Magán-Carrión et al. [9], I2P’s network database (netDB) which is based on the Kademlia algorithm uses a distributed hash table managed by floodfill routers that encrypts key records such as lease sets and routerinfo. This approach limits direct access to routing details [10]. A router attains this critical Floodfill status through one of two means: either by explicit manual configuration or by being automatically promoted by the network based on its sustained performance and reliability.

This study addresses a key knowledge gap by providing empirical data on network-layer interactions, explicitly focusing on nodes involved in tunnel formation. It utilizes data from I2P routers established over a prolonged observation to track the construction of tunnels, the frequency of node connections, and the evolution of routing across more than 50,000 nodes. The result opens new avenues for studying the garlic network’s structure, tunnel efficiency, node influence, and overlay network behavior or anonymous communications network (ACN) behavior. The discovery helps researchers understand secure communication and cyber threat modeling [11,12].

Beyond descriptive observation, this work advances I2P research through two primary contributions. First, we introduce SWARM-I2P, a novel framework for large-scale, controlled router deployment that enables reproducible experiments on tunnel formation and node selection capabilities not supported by earlier methodologies. Second, we release a network layer dataset of unprecedented scope that permits rigorous analysis of node selection dynamics and reveals the emergence of high-capacity nodes that function as a reputation-based backbone of the network.

# DATA DESCRIPTION

The study utilizes empirical data from live tunnel formations to examine node selection patterns in the Invisible Internet Project (I2P), also known as the "garlic network." We deployed Docker-containerized I2P routers configured to log interactions, automatically capturing node selection dynamics over time. The dataset presented in Table 1 documents the discovery and routing behavior observed in the I2P network layer.

**Table 1**

**Dataset Files and Their Content with Explanation**

| File Name | Content |
|---|---|
| 1-Client-Tunnel.csv<br>2-Client-Tunnel-anonymized.csv | Data about the tunnels nodes used by clients to connect to the network. |
| 3-Fastset-Nodes.txt<br>4-Fastset-Nodes-By-Time.txt<br>5-Fastset-Frequency.txt | Information on "FastSet" nodes, including their characteristics and how frequently they appear or are used over time. |
| 6-High-Capacity-Set.txt<br>7-High-Capacity-Set-Freq.txt<br>8-High-Capaity-Freq2.txt | Data on nodes with high bandwidth and capacity, and the frequency of these nodes. |
| 9-Profile-By-Country.csv<br>10-Profiles-By-Country-anonymized.csv | Geographic distribution data showing the location of I2P nodes by country. |
| 11-FF-ClientTunnel.txt<br>12-FF-ExploratoryTunnel.txt<br>13-FF-ParticipatingTunnel.txt | Data related explicitly to different tunnel types (client, exploratory, and participating) for Floodfill routers. |
| 14-nonFF-ClientTunnel.txt<br>15-nonFF-ExploratoryTunnel.txt | Tunnel data for non-Floodfill (regular user router). |
| 16-Nodes-Multiple-Tunnels.csv<br>17-Nodes-In-MultiTunnels.csv<br>18-Nodes-In-MultiTunnels2.csv | The datasets focus on nodes that are part of multiple tunnels simultaneously. |
| 19-Participatory-Tunnel-anonymized.csv<br>20-Participatory-Tunnel2-anonymized.csv | Data on participatory tunnels (tunnels contributing to the network function). |
| 21-TrafficMetadata-anonymized.csv<br>22-TrafficMetadata_Anonymized.csv | Metadata about network traffic, including packet and byte counts. |
| 23-Exploratory-Tunnel.csv | Data collected from "exploratory" tunnels, which are tunnels used to map the network. |

**Tunnel and Selection Data**

The dataset comprises multiple text files and CSV files. In the file (1-Client Tunnel.csv), we have records of 340 nodes. The file (2-Client-Tunnel.csv) records I2P tunnel nodes with timestamps as shown in Table 2. It includes traffic direction, tunnel expiration, data volume, and node roles (gateway, participant, or endpoint). Each entry lists a node's abbreviated and full router ID, country, and IP address. Most nodes are in Germany, France, Sweden, and the Netherlands. This dataset reveals how client tunnels form and change over time. Tracking repeated nodes highlights geographic trends in I2P routing. Nodes that recur frequently are observed to be the preferred choice for tunnels, a pattern that points to the existence of a central core of popular nodes on the network. Our data enables a detailed study of real-world node behavior and tunnel dynamics. As shown in Fig 2, the top 20 most frequently selected nodes account for 4.62% of tunnel nodes, an observation that illustrates a concentrated distribution within the network.

**Table 2**

**Data Dictionary for Client Tunnel Activity (e.g., 2-Client-Tunnel-anonymized.csv)**

| Header Name | Data Type | Description |
|---|---|---|
| Timestamp | DateTime | Time the tunnel record was collected. |
| Direction | Text | The flow of traffic (Inbound or Outbound). |
| Expiration | Time/Int | The remaining time until the tunnel is scheduled to be torn down. |
| Data_Volume | Int (KiB) | The volume of data handled by the node in this tunnel session (Kibibytes). |
| Node_Role | Text | The function of the node in the tunnel (Gateway, Participant, or Endpoint). |
| Router_ID | Text | Unique identifier for the I2P router. |
| Country | Text | The detected geographic location of the node. |

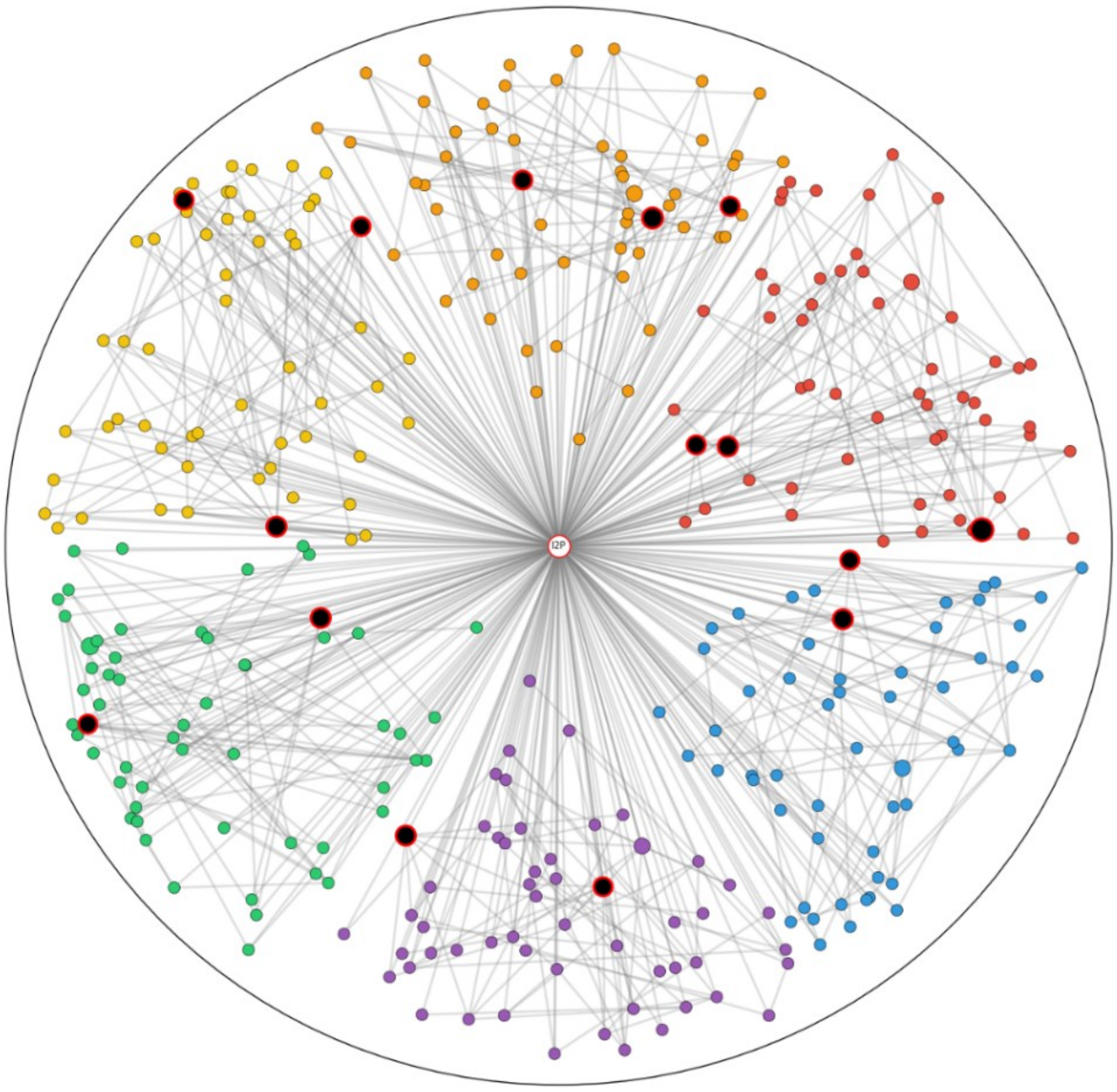

Fig 2. I2P Client Tunnel Node Visualization from a User router (Hyperbolic Clustered View) [Reproducibility Note: Data sourced from Zenodo file: Client-Tunnel.csv. Analysis script located in the GitHub repository at https://github.com/abksiddique/swarmi2p/blob/main/notebooks/I2P_Client_Tunnel_Node_Visualization.ipynb]

I2P tunnels serve three primary functions: (1) client tunnels for actual communications, (2) exploratory tunnels for network discovery and profiling, and (3) participatory tunnels for hub involvement [13]. Fig 3 illustrates these tunnel types as displayed in an I2P console interface of a fully immersed router we deployed on the Garlic Network. A hyperbolic disc view of the exploratory tunnel in Fig 4 reveals more nodes of different reputations discovered by profiling to help build tunnels.

e 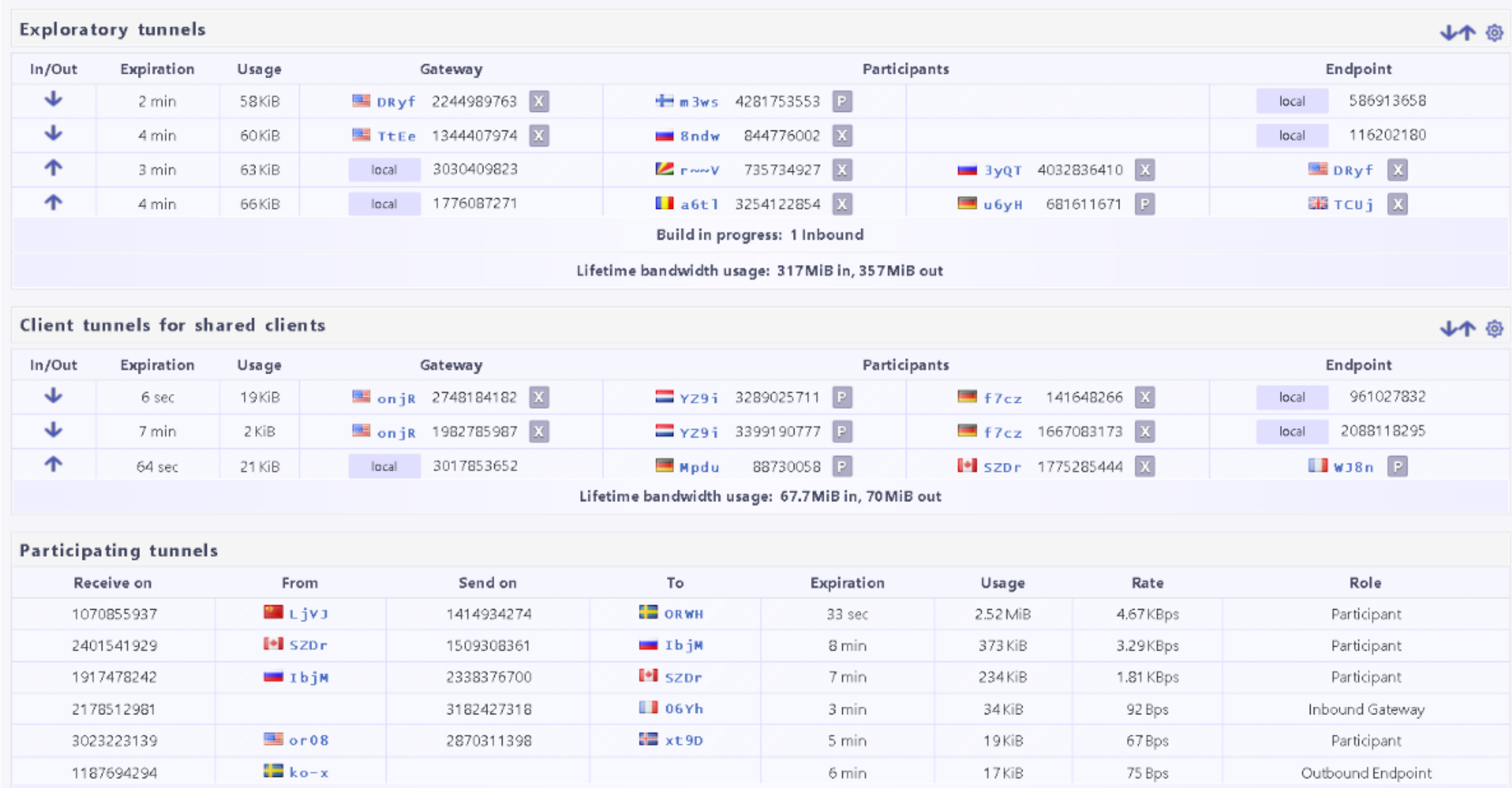

**Exploratory tunnels**

| In/Out | Expiration | Usage | Gateway | Participants | | Endpoint |
|---|---|---|---|---|---|---|
| ↓ | 2 min | 58KiB | DRyf 2244989763 X | m3ws 4281753553 P | | local 586913658 |
| ↓ | 4 min | 60KiB | TtEe 1344407974 X | 8ndw 844776002 X | | local 116202180 |
| ↑ | 3 min | 63KiB | local 3030409823 | r~~V 735734927 X | 3yQT 4032836410 X | DRyf X |
| ↑ | 4 min | 66KiB | local 1776087271 | a6tl 3254122854 X | u6yH 681611671 P | TCUj X |

Build in progress: 1 Inbound

Lifetime bandwidth usage: 317MiB in, 357MiB out

**Client tunnels for shared clients**

| In/Out | Expiration | Usage | Gateway | Participants | | Endpoint |
|---|---|---|---|---|---|---|
| ↓ | 6 sec | 19KiB | onjR 2748184182 X | YZ9i 3289025711 P | f7cz 141648266 X | local 961027832 |
| ↓ | 7 min | 2KiB | onjR 1982785987 X | YZ9i 3399190777 P | f7cz 1667083173 X | local 2088118295 |
| ↑ | 64 sec | 21KiB | local 3017853652 | Mpdu 88730058 P | SZDr 1775285444 X | WJ8n P |

Lifetime bandwidth usage: 67.7MiB in, 70MiB out

**Participating tunnels**

| Receive on | From | Send on | To | Expiration | Usage | Rate | Role |
|---|---|---|---|---|---|---|---|
| 1070855937 | LjvJ | 1414934274 | ORWH | 33 sec | 2.52MiB | 4.67KBps | Participant |
| 2401541929 | SZDr | 1509308361 | IbjM | 8 min | 373KiB | 3.29KBps | Participant |
| 1917478242 | IbjM | 2338376700 | SZDr | 7 min | 234KiB | 1.81KBps | Participant |
| 2178512981 | | 3182427318 | 06Yh | 3 min | 34KiB | 92Bps | Inbound Gateway |
| 3023223139 | or08 | 2870311398 | xt9D | 5 min | 19KiB | 67Bps | Participant |
| 1187694294 | ko-x | | | 6 min | 17KiB | 75Bps | Outbound Endpoint |

Fig 3. I2P tunnel Types: Exploratory, Client, and Participatory - User Console

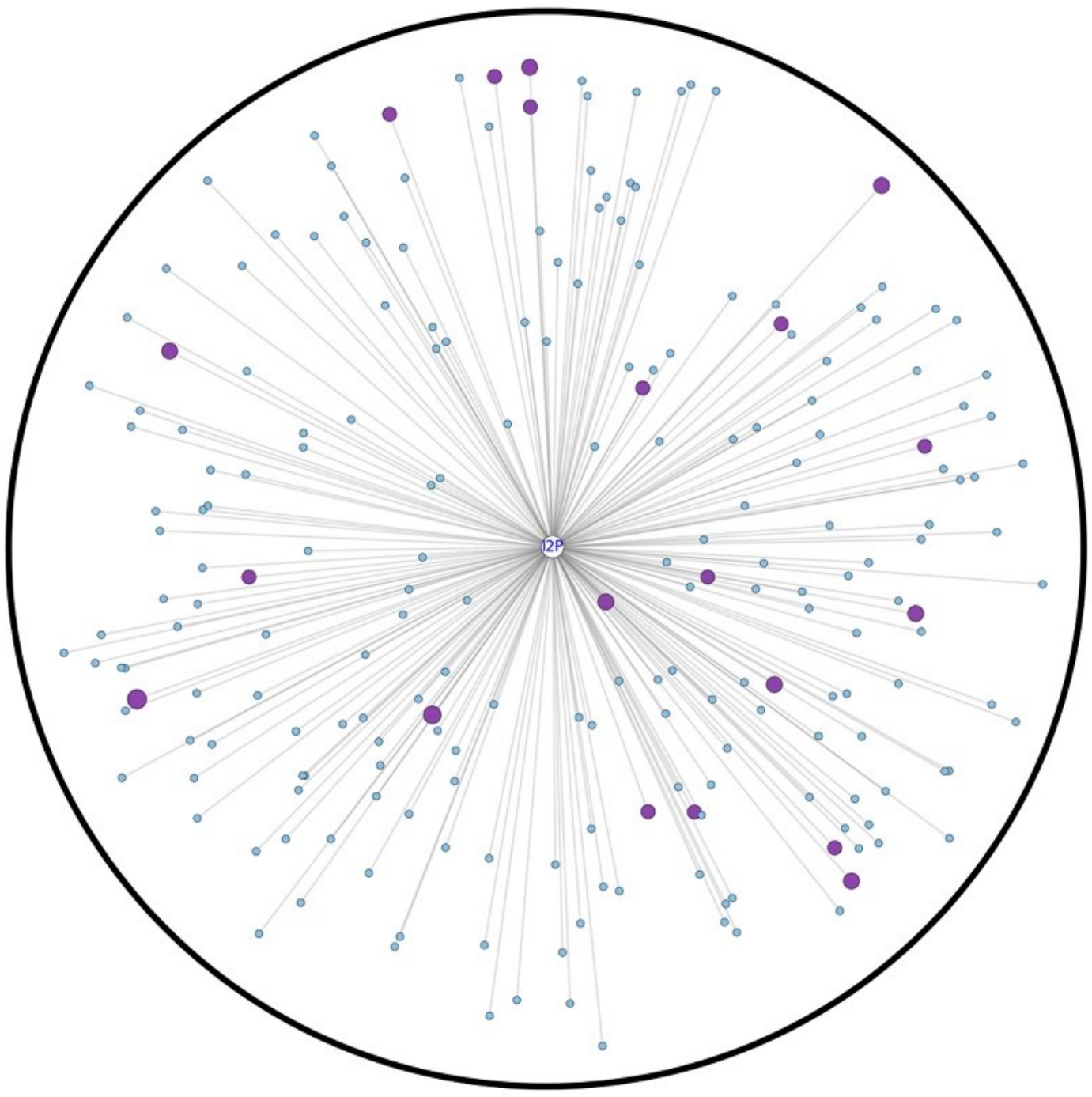

Fig 4. I2P Users Exploratory Tunnels nodes Discovery

[Reproducibility Note: Data sourced from Github file: [https://github.com/abksiddique/swarmi2p/blob/main/data/Exploratory%20Tunnel-merged.txt]. Analysis script located in the GitHub repository at https://github.com/abksiddique/swarmi2p/blob/main/notebooks/Exploratory_Tunnels.ipynb]

In the file (3-Fastset-Nodes.txt), we have 2,077 unique Fastset nodes identified through continuous profiling. This was achieved after our router was fully integrated into the Garlic network. Routers assessed nodes to determine if they could be part of their tunnel based on reputation metrics, including bandwidth, latency, uptime, and reliability. With few nodes meeting the criteria and categorized as fastset, signifying their suitability for stable, high-performance client communication tunnels (see Fig 5). Similarly, Fig 5 also presents the top 20 most frequently selected fastset nodes based on tunnel profiling using (4-Fastset-Nodes-By-Time.txt) and augmented by large-scale users' profiles in (5-Fastset-Frequency.txt) data.

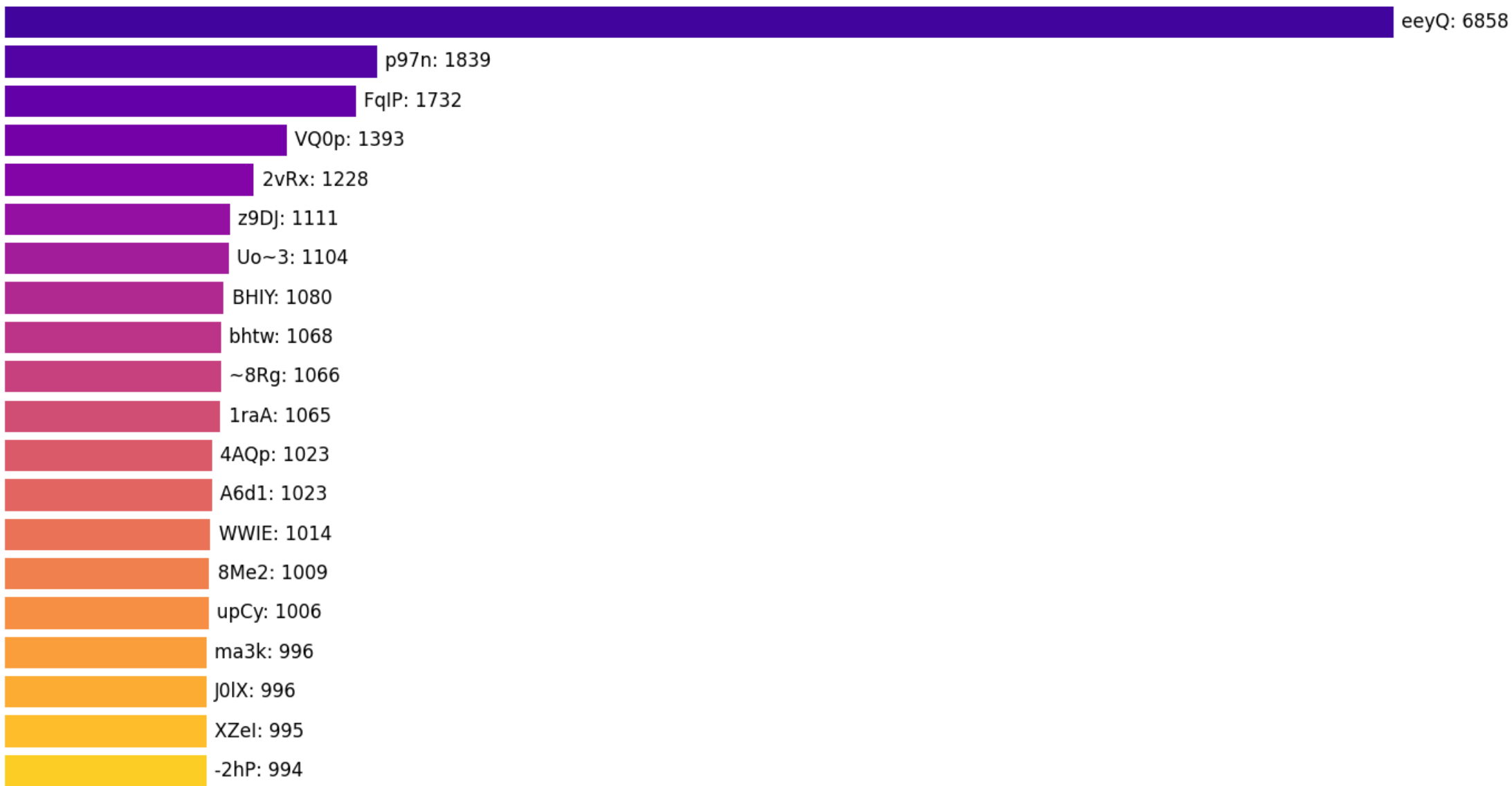

Fig 5. I2P Users' Client Tunnel Nodes

[Reproducibility Note: Data sourced from Zenodo file: 4-Fastset-Nodes-By-Time.txt. Analysis script located in the GitHub repository at https://github.com/abksiddique/swarmi2p/blob/main/notebooks/Fastset_Nodes.ipynb]

Data in (6-High-Capacity-Set.txt) and (7-High-Capacity-Set-Freq.txt) tracks over 2,331 high-capacity I2P nodes, which are critical for tunnel reliability. The median node appears 76 times. This indicates a core group of stable router nodes used in any user's exploratory tunnels or as a fallback for client tunnels. The top 20 nodes (e.g., Z1bh-f14..., appearing 3,926 times) are frequently selected for their high bandwidth and uptime, which indicates a preference for stable and high-performance nodes in tunnel formation. Fig 6 and Table 3 show the hierarchy where a fraction of high-performance nodes handles most tunnel traffic. Tunnels are recreated every few minutes (about 10 minutes) for security reasons [2] [14].

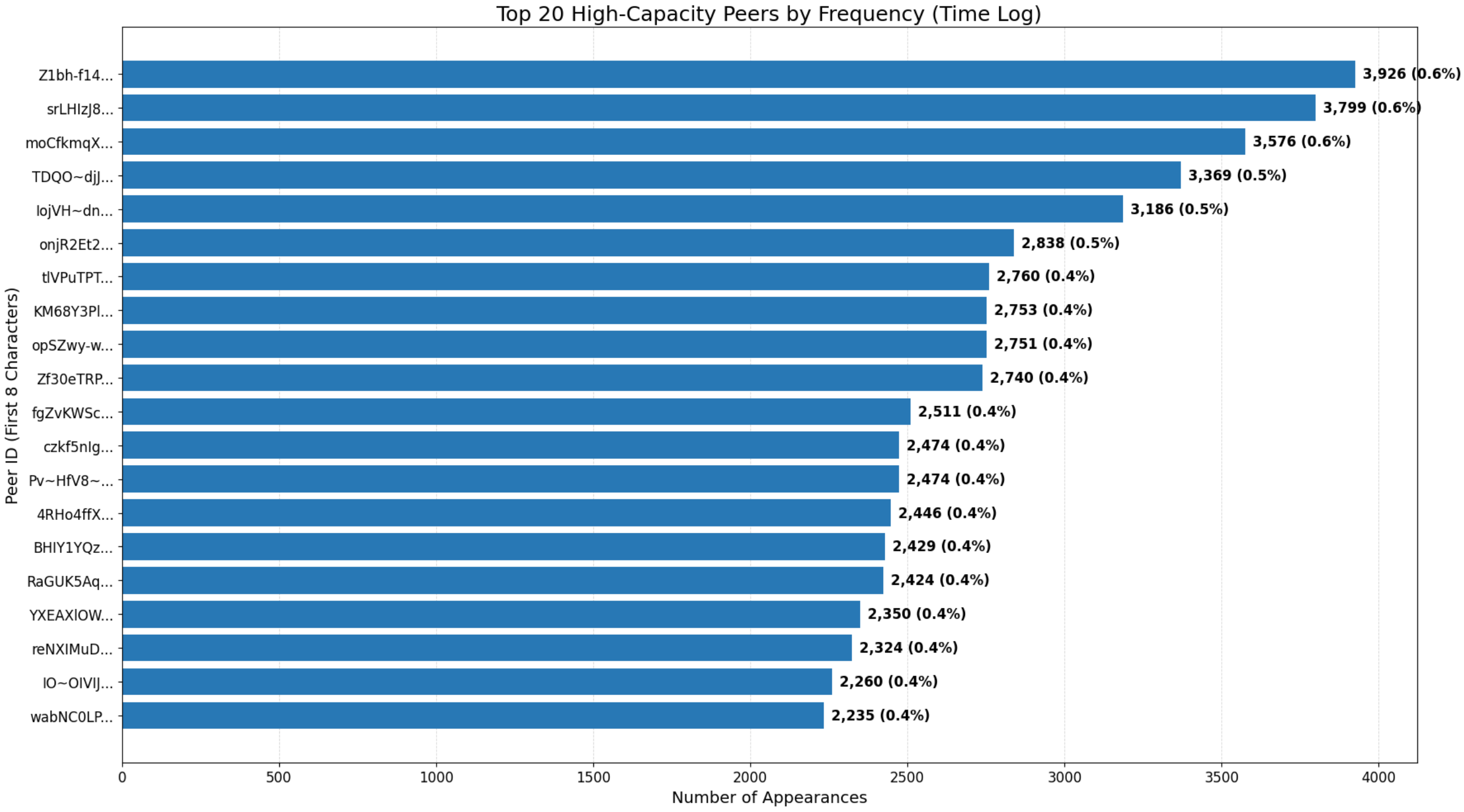

Fig 6. Top 20 High Capacity Node Profile

[Reproducibility Note: Data sourced from Zenodo file: (6-High-Capacity-Set.txt and 7-High-Capacity-Set-Freq.txt). Analysis script located in the GitHub repository at https://github.com/abksiddique/swarmi2p/blob/main/notebooks/High_Capacity.ipynb]

**Table 3**

**Top 20 I2P nodes by selection frequency**

| **Node ID (First 8 chars)** | **Frequency** | **% of Total** | **Cumulative %** |
|---|---|---|---|
| **TU4AVivT...** | 687 | 12.2 | 12.2 |
| **hA~tqWxf...** | 585 | 10.39 | 22.58 |
| **N7T9mVbu...** | 425 | 7.54 | 30.13 |
| **x-eBfsgO...** | 417 | 7.4 | 37.53 |
| **aTdMtnxQ...** | 395 | 7.01 | 44.54 |
| **WwNGsm99...** | 309 | 5.49 | 50.03 |
| **dz3ON-1U...** | 301 | 5.34 | 55.37 |
| **ZDz09qEx...** | 270 | 4.79 | 60.16 |
| **Fu-8GZY8...** | 260 | 4.62 | 64.78 |
| **IwOkup7K...** | 255 | 4.53 | 69.31 |
| **g5xHgSjt...** | 221 | 3.92 | 73.23 |
| **IiN-TYoM...** | 219 | 3.89 | 77.12 |
| **twbWgUGo...** | 196 | 3.48 | 80.6 |
| **CpMHp1xj...** | 178 | 3.16 | 83.76 |
| **3UavGAjy...** | 174 | 3.09 | 86.85 |
| **~Go801IA...** | 168 | 2.98 | 89.83 |
| **ZB~nvHJj...** | 151 | 2.68 | 92.51 |
| **SLIdpSD6...** | 144 | 2.56 | 95.06 |
| **F9577L-s...** | 140 | 2.49 | 97.55 |
| **bUJ0DwK8...** | 138 | 2.45 | 100 |

| timestamp | Gini | entropy | share_XfR |
|---|---|---|---|
| 2024-11-17 10:00:09 | 0.671497 | 8.68676 | 0.0967742 |
| 2024-11-17 10:30:09 | 0.670402 | 8.71162 | 0.09375 |
| 2024-11-17 11:00:09 | 0.66527 | 8.74944 | 0.0909091 |
| 2024-11-17 11:30:09 | 0.667267 | 8.78537 | 0.0882353 |
| 2024-11-17 12:00:09 | 0.668076 | 8.82306 | 0.0909091 |
| 2024-11-17 12:30:09 | 0.672325 | 8.79891 | 0.0882353 |
| 2024-11-17 13:00:09 | 0.66841 | 8.78667 | 0.0909091 |
| 2024-11-17 13:30:09 | 0.669235 | 8.78671 | 0.09375 |
| 2024-11-17 14:00:09 | 0.675977 | 8.76115 | 0.0909091 |
| 2024-11-17 14:30:09 | 0.668072 | 8.78364 | 0.09375 |

**Node Profile and Geographic Data**

Similarly, in (8-High-Capaity-Freq2.txt) data, the number of times a high capacity is used shows the most critical nodes in the network. The data in (9-Profiles-By-Country.csv and 10-Profiles-By-Country.csv) comprises over 3,444 records. Each contains a timestamp, unique node identifier, country, group classification, capability code (Caps) as detailed in Table 4, I2P software version, speed, and capacity.

**Table 4**

**I2P Router "Caps" field and Explanations**

| Capability Flag (Caps) | Meaning |
|---|---|
| D | Congested |
| F | Floodfill |
| K | Under 12 KBps shared bandwidth |
| M | 49 - 65 KBps shared bandwidth |
| O | 131 - 261 KBps shared bandwidth |
| R | Reachable |
| X | Over 2048 KBps shared bandwidth |
| E | Severely congested |
| G | Not accepting tunnels |
| L | 12 - 48 KBps shared bandwidth |
| N | 66 - 130 KBps shared bandwidth |
| P | 262 - 2047 KBps shared bandwidth |
| U | Unreachable |

The visualization in Fig 7 offers an overview of data by profile country and is supported by a data dictionary in Table 5. The bar chart on the top left shows that the United States leads the top 10 countries in terms of node count, indicating a concentration of activity in a few regions. The histogram illustrates the distribution of speed and capacity, with speed exhibiting a mean of 121.21 and a standard deviation of 48.50, and capacity showing a mean of 8.57 at a standard deviation of 1.20. The scatter plot visualizes a trend where nodes with higher capacity appear to achieve higher speeds, indicating a relationship between these two metrics.

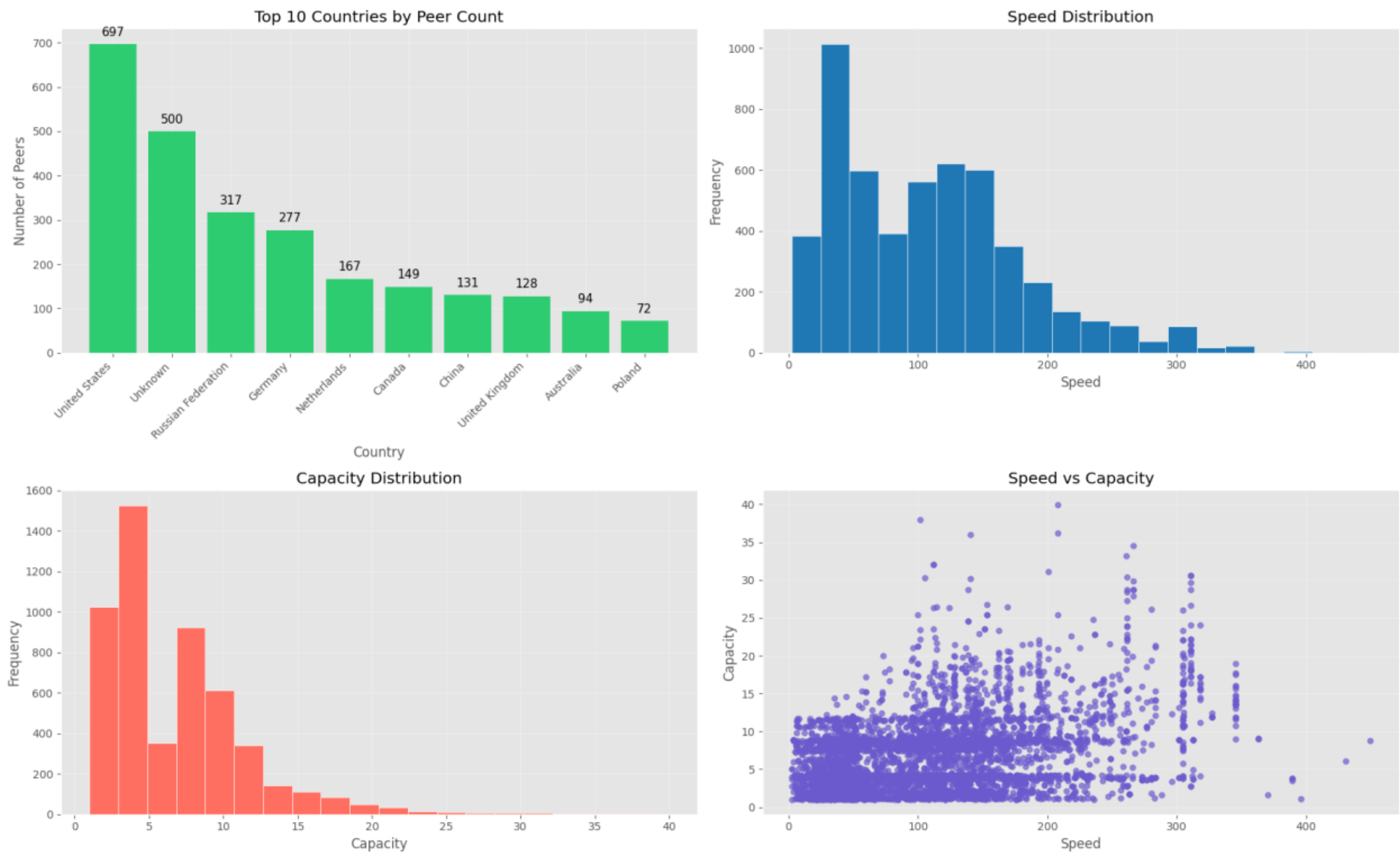

Fig 7. I2P Node Profile with Country, Capacity, Speed

[Reproducibility Note: Data sourced from Zenodo file: (9-Profiles-By-Country.csv and 10-Profiles-By-Country.csv). Analysis script located in the GitHub repository at https://github.com/abksiddique/swarmi2p/blob/main/notebooks/Profiles_Country.ipynb]

**Table 5**

**Data Dictionary for node profile (e.g., 9-Profiles-By-Country.csv)**

| Header Name | Data Type | Description |
|---|---|---|
| Country | Text | The detected geographic location of the node. |
| Node Abbrev | Text | Abbreviated (short) identifier for the I2P router. |
| Full Node ID | Text | The unique, full identifier for the I2P router. |
| Groups | Text | Classification based on network role (e.g., standard or floodfill). |
| Caps | Text | Capability Flag detailing node functionality and capacity (refer to Table 4). |
| Version | Text | The I2P router software version. |
| Speed | Float/Int | Measured network speed of the node. |
| Capacity | Float/Int | Measured routing capacity of the node. |
| Integration | Text | Status of the router's integration or participation. |
| Status | Text | Operational status (e.g., Reachable, Unreachable). |
| IP Address | Anonymized Text | IP address of the node. |

The ”Caps” field provides critical insights into router capabilities through standardized flags as detailed in Table 4, where ”X” designates high-bandwidth nodes (2048 KBps), ”f indicates floodfill functionality, and ”R” confirms reachability status [15]. These markers enable precise analysis of node contributions to network performance, with combinations like ”XfR” identifying high-capacity and reachable floodfill routers. Our dataset captures this operational diversity through categorized logs contrasting floodfill and standard router behaviors. The floodfill records (11-FF-ClientTunnel.txt, 12-FF-ExploratoryTunnel.txt, and 13-FF ParticipatingTunnel.txt) comprehensively capture client traffic, network probing results, and routing participation. This reflects the floodfill routers’ dual role in both data transmission and network maintenance (via the netDb). In comparison, the non-floodfill data (14-nonFF-ClientTunnel.txt and 15-nonFF ExploratoryTunnel.txt) reveal the more limited but equally crucial functions of standard routers as nodes. This structural comparison illustrates how floodfill routers, in fulfilling their database maintenance duties, observe significantly more nodes than their standard counterparts.

### Multi-Tunnel and Traffic Metadata

The file (16-Nodes-Multiple-Tunnels.csv) tracks I2P router nodes simultaneously appearing in several tunnels. This data helps us understand how the nodes contribute to the network’s routing structure. Each entry displays the date and time, their anonymous ID, the country they appear to be in, the number of tunnels they joined, and the amount of data they handled measured in kibibyte (KiB). The six charts in Fig 8 reveal several essential patterns. The United States has the most active nodes, with Russia, Germany, and Canada also showing strong participation. The charts illustrate data usage by country, with the average usage of nodes in Russia and Canada appearing higher than that of other nations.

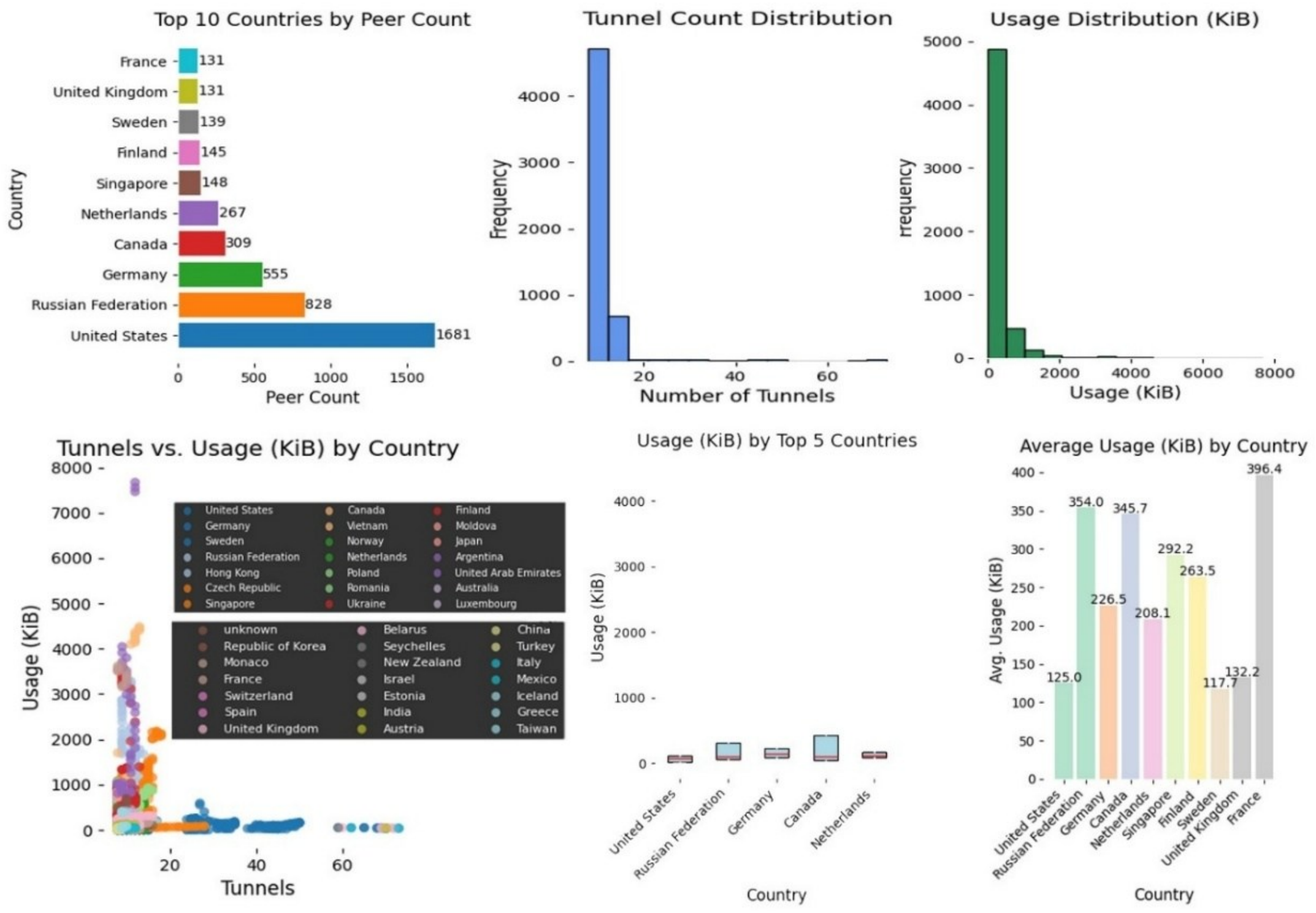

Fig 8. I2P Visualization of nodes that appear in multiple tunnels

[Reproducibility Note: Data sourced from Zenodo file: (16-Nodes-Multiple-Tunnels.csv). Analysis script located in the GitHub repository at https://github.com/abksiddique/swarmi2p/blob/main/notebooks/Peers_Multiple_Tunnels.ipynb]

The data in (17-Nodes-In-MultiTunnels.csv) and (18-Nodes-In MultiTunnels2.csv), visualized in Fig 9, illustrates key network traffic patterns. Bar charts display data volume for Top Talkers (Bytes Sent) and Top Destinations (Bytes Received). Among these, IP node (10.211.1.xxx) shows high transmission activity, while (176.98.182.xxx) receives significantly fewer bytes. Network graphs (A, B, C) depict communication structures: graph (A) reveals a star topology with node A as the central hub, while others show mesh configurations. The traffic overtime (1-minute resolution) graph highlights fluctuating traffic, peaking near 10 bytes and averaging around 6 bytes. A heatmap further visualizes interaction intensity between source-destination IPs, with darker shades indicating higher traffic frequency, particularly involving (10.211.1.xxx).

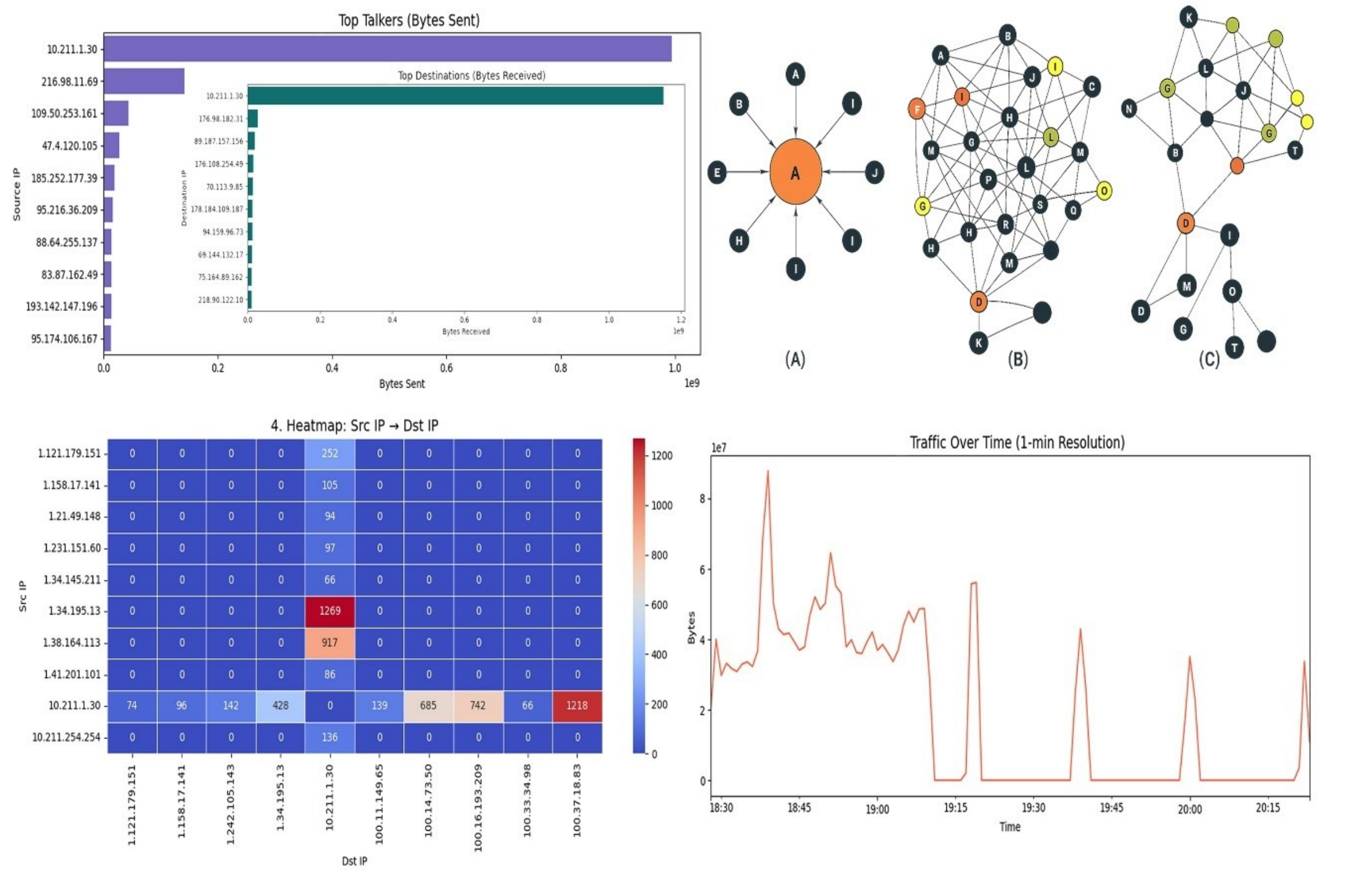

Fig 9. Network Traffic pattern for nodes in Multiple Tunnels

[Reproducibility Note: Data sourced from Zenodo file: (17-Nodes-In-MultiTunnels.csv) and (18-Nodes-In MultiTunnels2.csv). Analysis script located in the GitHub repository at https://github.com/abksiddique/swarmi2p/blob/main/notebooks/Nodes_MultiTunnels.ipynb]

Detailed metadata on I2P tunnel activities on client, participating, and exploratory tunnels, has also been captured. The file (19-Participatory Tunnel.csv) records node roles, session direction, expiration, and IP information, offering insight into how nodes forward traffic. Logs in (20-Participatory Tunnel2.csv) help study relay behavior by showing data usage, timing, and stability for intermediary nodes. A broader set of tunnel traffic metadata, including protocol types, data sizes, and IP flows, is included in files (21-TrafficMetadata.csv and 22-TrafficMetadata.csv) as shown in Table 6, supporting the analysis of communication volume and node interaction. The total traffic metadata records amount

to 1,997 records (1,003,032 bytes) and 4,222,793 records (2,147,585,625 bytes). Additionally, (23-Exploratory-Tunnel.csv) captures the exploratory tunnel sessions over a 24-hour period, demonstrating how nodes test and build paths across the network.

**Table 6**

**Data Dictionary for traffic metadata (e.g., 21-TrafficMetadata.csv)**

| Time | DateTime | Time the traffic record was captured. |
|---|---|---|
| Src IP | Anonymized Text | Anonymized source IP address. |
| Src Port | Int | Source port number. |
| Dst IP | Anonymized Text | Anonymized destination IP address. |
| Dst Port | Int | Destination port number. |
| Proto | Text | Network protocol type (e.g., RTCP, UDP, SSU). |
| TTL | Int | Time To Live (Network hop count). |
| PacketSize(Bytes) | Int | The size of the packet or traffic event in bytes. |

Together, these datasets reflect the dynamic and decentralized nature of I2P, supporting fine-grained analysis of routing, node roles, and geographic patterns. We have prepared this data carefully for researchers. The files utilize standard formatting compatible with Python tools, such as Pandas, or statistical programs like R, and are clearly labeled and well-documented. Most importantly, we removed identifying information from the datasets to protect users' privacy. The dataset gives researchers unique insight into I2P's operations without compromising privacy protections. Scientists can use it to study how traffic is routed, which countries have the most reliable nodes, and how the network balances its load. This is beneficial for cybersecurity experts working to improve anonymous networks and law enforcement agencies (LEAs) that are interested in understanding patterns.

## Node Selection Modeling

To analyze how tunnel selections reflect structural behavior in the I2P network, we applied a basic probabilistic model to the observed node selection frequencies. Let $fi$ denote the number of times node $i$ is selected in any tunnel within the set under consideration, let $n$ be the number of nodes in that set (e.g., $n$ =20 for the top-20 subset), and define the total selections as $T \ = \sum_{i=1}^{n} fi$ . The empirical probability of selecting a node $i$ is defined as:

$$P\ i = \frac{fi}{T} \qquad 1$$

This distribution allows us to study dispersion and inequality in the node selection process. To assess the dispersion of node selection, we compute Shannon entropy (base-2; bits) as:

$$H = -\ \sum_{i=1}^{n} pi\ \log_2 pi \qquad 2$$

Lower entropy values indicate that a few nodes dominate tunnel participation, while higher values suggest a more balanced selection across the node set. To quantify inequality directly, we compute the population Gini coefficient $G$ over the same set (with $n$ nodes and total $T$) as:

$$G = \frac{\sum_{i=1}^{n}\sum_{j=1}^{n}|f_i - f_j|}{2n\sum_{i=1}^{n} f_i} \qquad 3$$

Using (1)– (3), we quantify dispersion and inequality in node selection. For reference, G ranges from 0 (even selection) to 1 (maximal inequality). Applying these metrics to the top 20 most frequently selected High-Capacity nodes yields Shannon entropy H=4.15 bits and Gini coefficient G=0.27. This moderate Gini value (≈0.27) indicates that selection within this elite subset is predictably distributed rather than perfectly equal, demonstrating a moderate level of concentration. In the I2P setting, this alignment with capacity-aware selection means that once routers qualify as high-capacity, their selection probabilities within this elite set are similar, even though, across the whole network (>50,000 nodes), overall selection remains concentrated in a comparatively small fraction of nodes.

To evaluate short-term stability and deepen the analysis, we extended the entropy and Gini measures to rolling two-hour windows (30-minute step) in the High-Capacity event stream. This temporal analysis revealed a significant empirical finding of high network concentration: the average Gini coefficient stabilized around 0.68 (Entropy≈8.73 bits). This high value, far from the ideal equality (G=0.0), suggests that selection within short operational windows is significantly more hierarchical and concentrated than assumed by the network's decentralized design goals. This finding empirically validates theoretical claims regarding operational bias in anonymity networks [16]. Capability shares were steady across windows, with XfR nodes consistently represented. Correlations between window metrics and capability shares are descriptive in short slices but become robust over longer traces.

# EXPERIMENTAL DESIGN, MATERIALS AND METHODS

The SWARM-I2P framework, as shown in Fig 10 is developed to empirically study the Invisible Internet Project (I2P) network through large-scale router deployment and traffic analysis. Our deployment architecture combined local computational resources at the University of Cincinnati, Ohio Cyber range (OCRI), with connection to cloud-based virtual private servers (VPS) running Linux Ubuntu 24.04 LTS. We used Uncomplicated Firewall (ufw) on cloud VPS to allow traffic via the system's port. Our hybrid approach enabled secured and controlled experimentation and real-world observation of the garlic network (I2P network).

The hybrid architecture, which combines local computational resources with a cloud-based VPS tunnel, was superior to a cloud-only deployment for several key reasons. This technique enabled us to utilize institutional compute facilities on a large scale for an extended period, which would have been cost-prohibitive in a purely commercial cloud environment. This setup also provided a secure means for our I2P routers to access the Garlic Network as agents, thereby mitigating potential security attacks and obstructions, as well as firewall restrictions, and prioritizing distributed access to the network. This ensured a controlled and reproducible environment for experimentation.

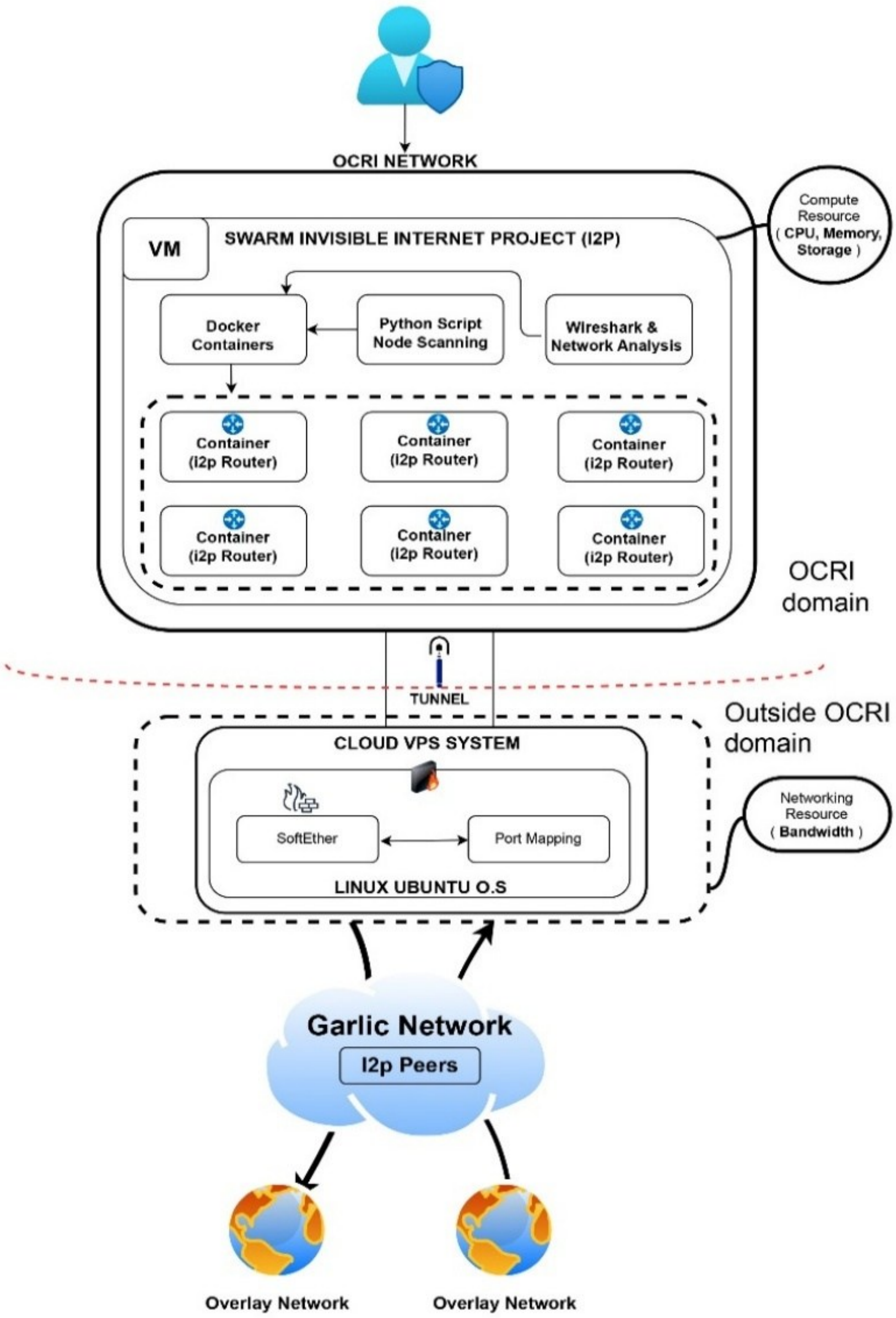

Fig 10. SWARM-I2P Framework for deploying I2P in large-scale

**Deployment Architecture and Dynamic Port Mapping**

Containerization formed the core of our deployment strategy. Using Docker with Portainer management interface, as shown in Fig 11, We instantiated multiple I2P router instances while overcoming port contention challenges through dynamic port mapping. The physical host allocated ports (30000-50000) were programmatically mapped to standard I2P local internal ports, such as 7657 for HTTP and 4444 for HTTP proxy for browsing [17]. We developed custom YAML (.yml) configurations, hosted on a GitHub project repository, to deploy the I2P routers [18]. On the cloud VPS, we designed a tunnel using SoftEther VPN server as shown in Fig 12 to establish a secure connection to the two systems using the developed framework in Fig 10. The tunnel helps mitigate firewall restrictions and institutional network limitations of OCRI, as well as effectively initiating each I2P router to the garlic network. This design enabled the simultaneous operation of hundreds of routers on a single hardware platform, a significant improvement over the limitations in prior studies, as reported in [15]. A router is initialized by executing the command (./i2prouter start) on the command line (terminal) [19]. The developed Dockerfile in [18] automatically generates the necessary i2p files for containerizations, such as (.i2p and i2p). Routers that act as floodfill routers were elevated by appending critical parameters to their “router.config” files using the configuration settings as found in Fig 13.

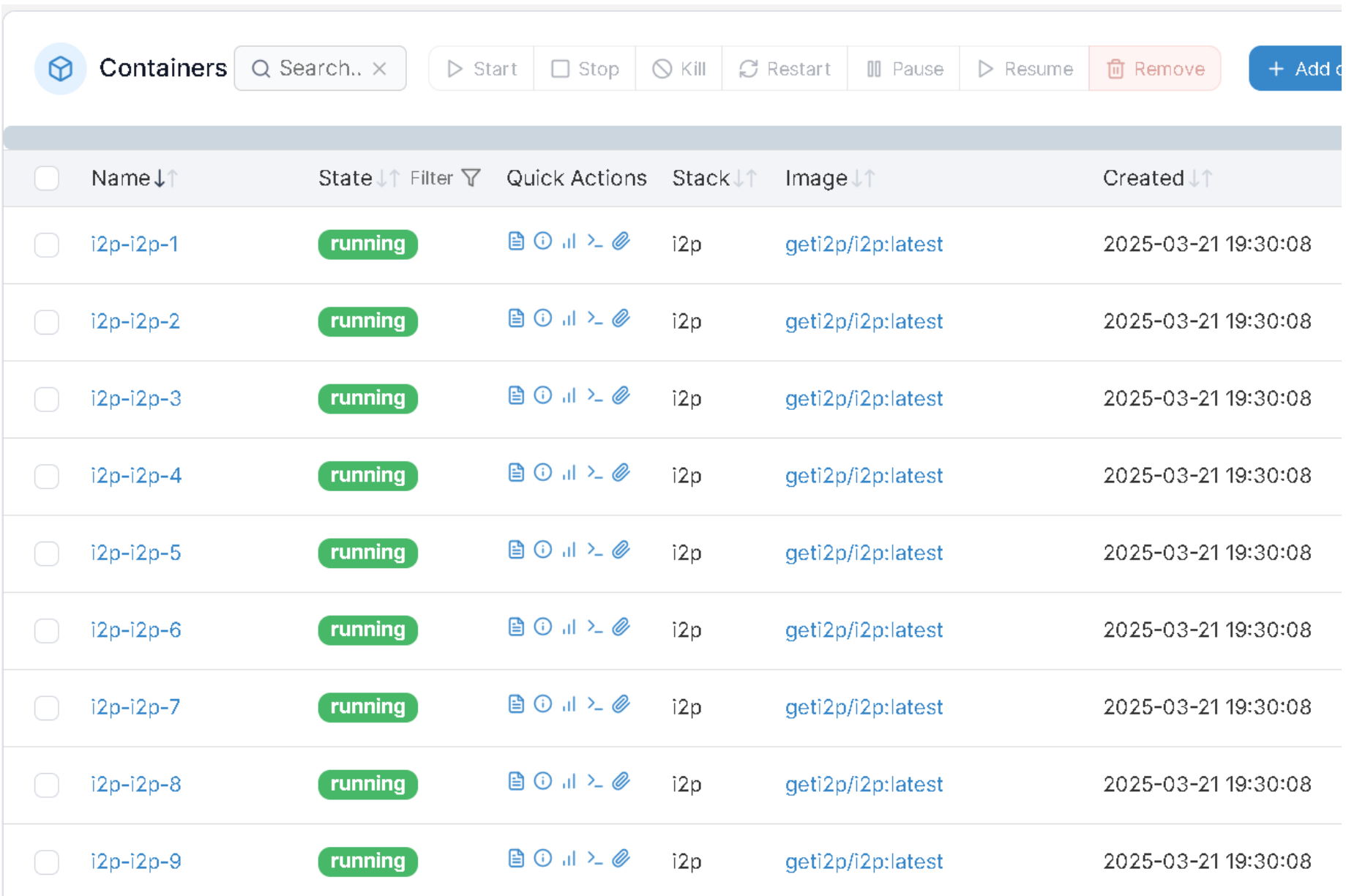

Fig 11. I2P Docker Container deployment on Portainer - Ubuntu

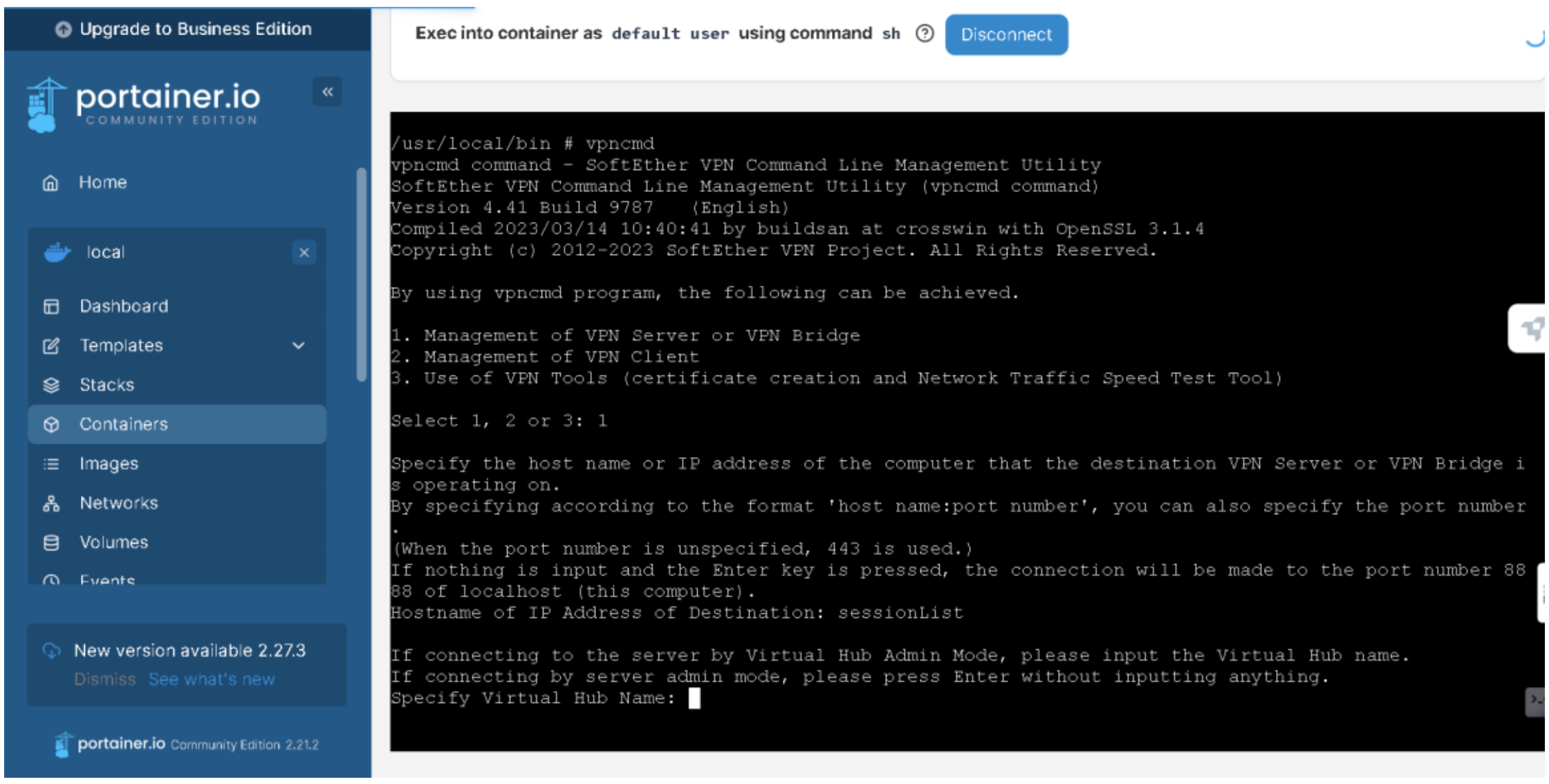

Fig 12. SoftEther Tunnel for Garlic Network Integration

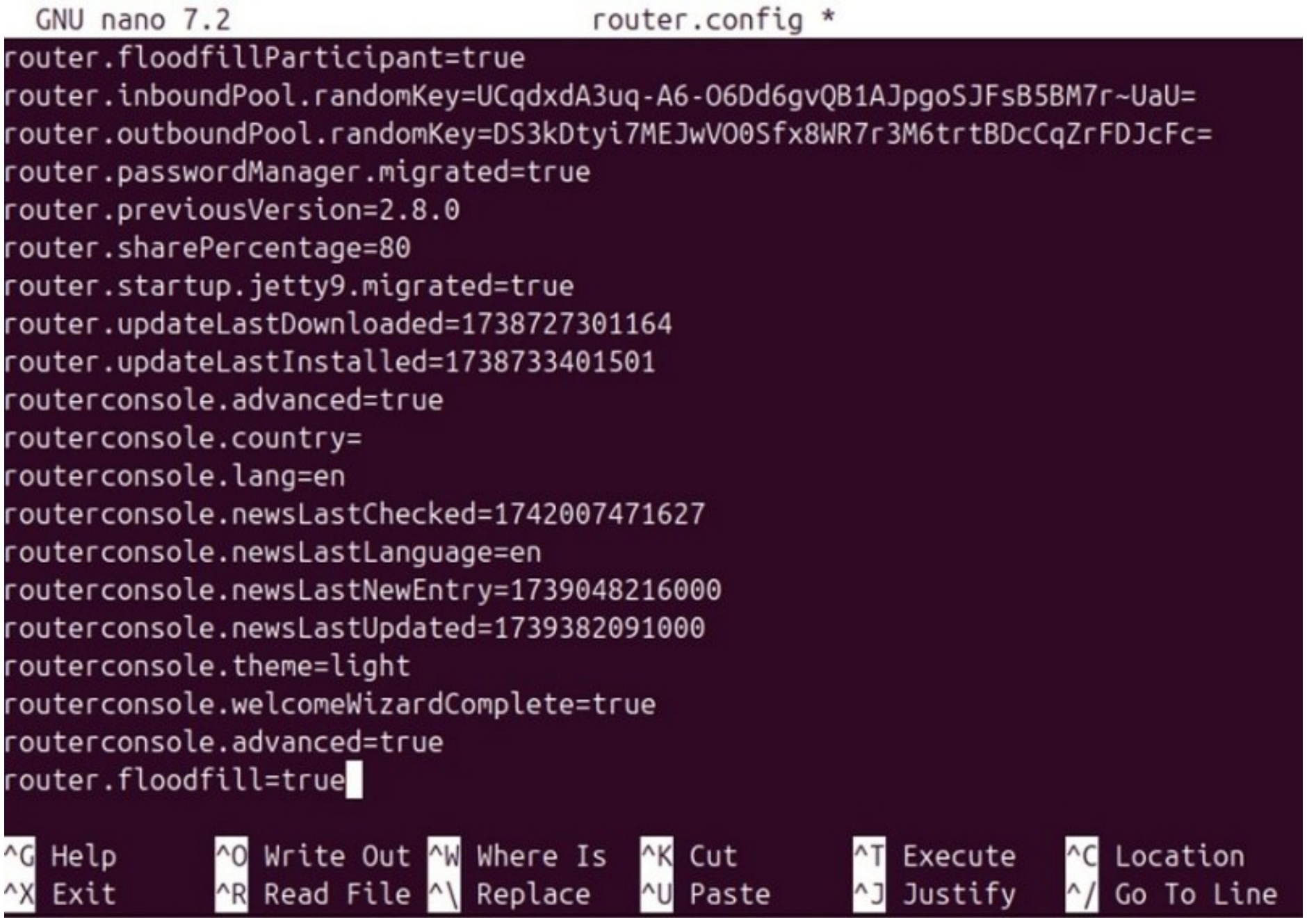

Fig 13. I2P router "config" file configuration

These floodfill nodes play a vital role in maintaining I2P's distributed network database (netDb), which implements the Kademlia algorithm for decentralized node coordination. Our technique enabled the observation of netDb operations and the construction patterns of tunnels. The SWARM-I2P framework utilizes a dedicated tunnel for controlled egress to the I2P network. This architecture ensures that, despite sharing a single external connection, the validity of node selection remains robust because I2P's tunnel construction is based on locally computed performance metrics and exclusive OR (XOR) distance in the netDb, rather than on external IP addresses or geographic location [16]. Routers rely on continuous node profiling derived from observed behavior to make these selection decisions. Our framework was designed to operate within this decentralized logic, making observations on node selection robust and valid regardless of the physical egress point. To support clarity and reproducibility, the architecture of the SWARM-I2P system is visually captured through two complementary diagrams in Fig 14. The class diagram and entity relationship (ER) diagram in Fig 14 illustrate the object-oriented structure of the system, where the SWARM-I2P component orchestrates the deployment of multiple I2P router instances. Each router maintains a set of tunnels; profiles observed nodes and optionally assumes floodfill roles after meeting the reputation mechanism. The DataCollector interacts with routers to log node activity and extract network-layer metadata. The entity-relationship diagram (Fig 14) formalizes the logical relationships between core entities in the system. It models the interaction between nodes in a tunnel. LogEntry reflects how router instances generate, store, and expose metadata over time. Each node is linked to multiple performance profiles and can construct or participate in several tunnel sessions. HostsEntry and DataCollection support structured access to router-level identifiers and traffic metrics.

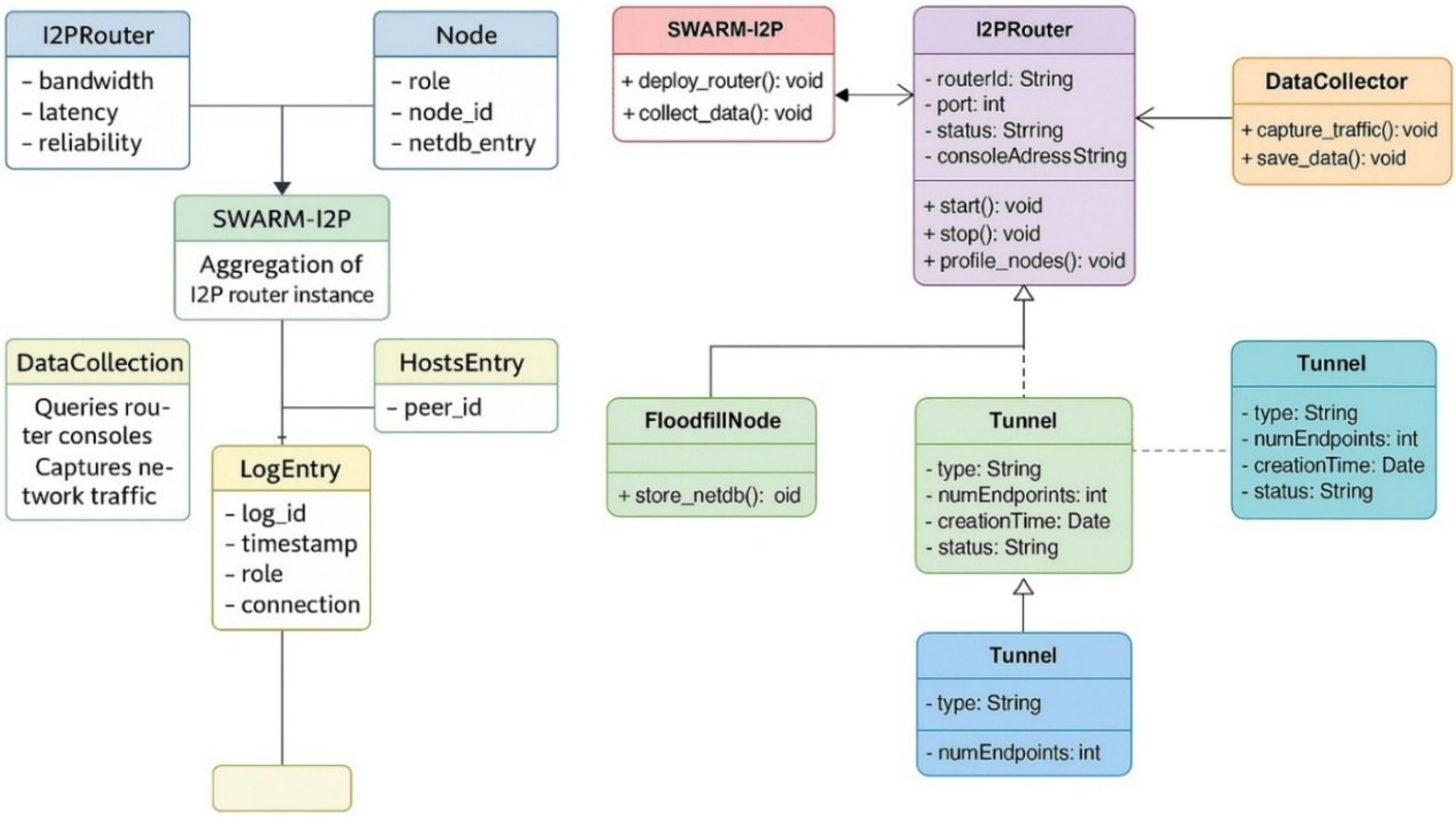

**Fig 14.** ER diagram on left and Class Diagram on right

**Router Swarm Synchronization**

To synchronize data across the router swarm, we automated the collection of tunnel information using Bash and Python scripts [18,20]. Following the method in Algorithms 1 and 2, these scripts polled each router's console (every minute) via the local endpoint 127.0.0.1:p/tunnels (where "p" is the port, for example, 40137). Each router was configured to log timestamped records of node interactions and tunnel assignments. The aggregation of these logs produced a comprehensive dataset, which we used to analyze temporal patterns in network behavior [18]. I2P's routing relies on frequent tunnel changes and high node churn, which necessitates a balanced approach to measurement. The 60-second synchronization interval was strategically engineered to function as a core strength of the SWARM-I2P framework. Although the nominal tunnel lifetime is established in literature to be approximately 10 minutes[14], I2P's security model mandates continuous tunnel recreation to prevent analysis of communication patterns, leading to high-frequency node churn. The 60-second polling interval was chosen to maximize the observation of these tunnel formation and node selection dynamics while strictly adhering to ethical compliance by minimizing the computational load imposed on node routers. This ensures long-term ecological validity and reproducibility of our observations.

We also utilize and examine access data of routers “hosts.txt” files within the i2p directory, as well as in (.i2p/netDb) to evaluate the router profiles organized alphabetically, as illustrated in Fig 15. Using wireshark and tcpdump, We captured network traffic with custom filters for I2P protocol analysis. As shown in Algorithms 1 and 2, we deployed a synchronized swarm of I2P routers to collect network layer topology data. I2P routers autonomously profile nodes based on direct communication history. They maintain internal reputation mechanism metrics such as latency, success rate, and reliability derived from observed behavior during message delivery and tunnel participation. While network database (netDb) provides router identity and advertised capabilities, node performance metrics are not obtained through netDb queries but are locally computed through empirical interactions of the nodes to build profiles. The system replicates I2P’s native tunnel selection: fast nodes for client tunnels and high capacity for exploratory tunnels, with optional third-party participation. Performance thresholds dynamically update node classifications (FastSet/High-Capacity) or otherwise. All routers log timestamped node interactions and tunnel assignments, generating an aggregated dataset that reveals temporal patterns in network behavior. The overall data collection workflow, from router deployment to final data curation, is visually summarized in Fig 16.

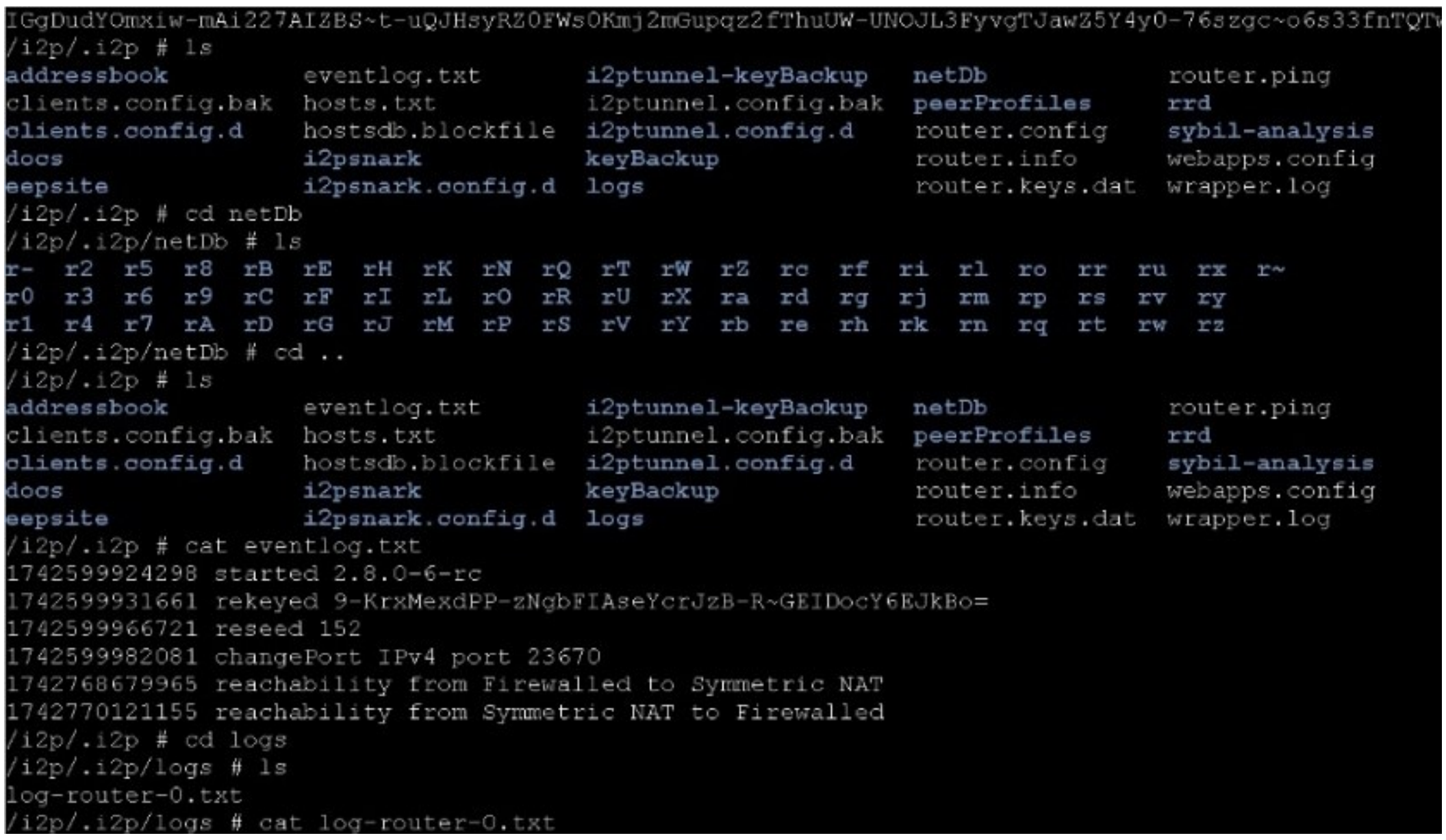

Fig 15. I2P Directory structure and files

**Algorithm 1 SWARM-I2P Deployment Procedure**

**procedure** Deploy SWARM-I2P
  Establish Tunnel, bridging the local VM and cloud VPS (overlay network)
  Launch N Docker containers for I2P routers (using Portainer for orchestration)
  Assign unique ports in the 30000–50000 range to each container's I2P instance
  In each container, initialize the I2P router by running (. /i2prouter start)
  **if** floodfill service is required (optional) **then**
    Elevate one or more I2P router instances to floodfill role
  **end if**
**end procedure**

**Algorithm 2 SWARM-I2P Data Collection Procedure**

**procedure** Collect SWARM-I2P Data
  **for** each I2P router container *r* **do**
    Query *r*'s web console at 127.0.0.1: p /tunnels for tunnel information
    Parse the console output to extract active tunnels, endpoints, and counts
    Read *r*'s hosts.txt and retrieve all node entries from r's netDb directory
    **for** each node entry in r's netDb **do**
      Extract node ID, role (e.g. floodfill or standard), and connection details
      Log the node ID, role, and connection info with current timestamp
    **end for**
  **end for**
  **in parallel:** capture network traffic on interface using Wireshark/tcpdump
  Save all captured data and logged node information to CSV/TXT files for analysis
**end procedure**

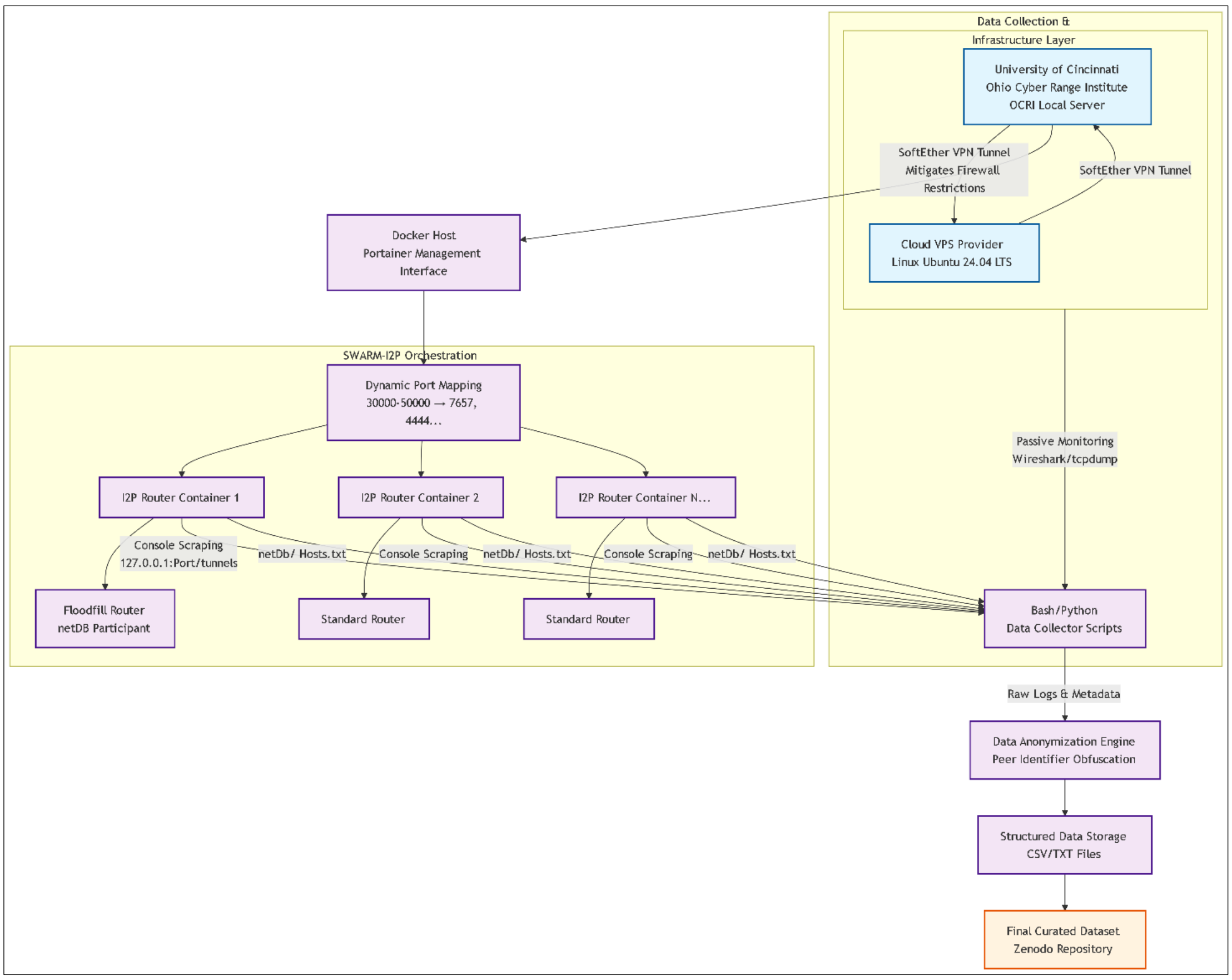

Fig 16. Flowchart as SWARM-I2P Data Collection Workflow

## Ethical Data Collection and Network Impact Mitigation

Our data collection methodology was rigorously designed to respect the core privacy principles of I2P network and maintain ethical compliance. We captured network traffic using by collecting only router-level metadata, such as node identifiers, tunnel timestamps, country location, and performance metrics. I2P's inherent security mechanisms, including garlic routing and secure transport (NTCP2/SSU), ensured that all captured data traffic comprised encrypted ciphertext. At no point was private user traffic intercepted, decrypted, or stored in plaintext, reinforcing that our analysis remained strictly focused on network-layer flows (e.g., packet counts and connection flows). To minimize our impact on the network, we carefully managed the rate of data collection to avoid disrupting normal operations. This cautious approach was vital for maintaining the study's validity and for operating ethically within a privacy-preserving system. Finally, as shown in Fig 16, to ensure both scientific utility and absolute user privacy, we employed a two-fold anonymization technique. The core node identifiers (Router Hashes) are inherently pseudonymous, as they form the immutable pseudonyms required for longitudinal analysis of network stability. Active anonymization was applied to network metadata, with specific network addresses masked through explicit truncation (e.g., replacing octets with 'xxxx'), thereby preventing any linkage to a particular user location.

# LIMITATIONS

While this study provides novel insights into I2P’s network layer, certain constraints merit acknowledgment. The dynamic nature of node participation means our data may reflect temporal network states rather than static snapshots, a fundamental characteristic of decentralized systems that our longitudinal design intentionally captured. The use of an egress point via the VPN could may potentially introduce a mild bias in peer discovery and selection toward that egress location. However, as discussed in our methodology, I2P's decentralized, hash-based routing and local peer profiling mitigate this potential bias, making our findings on selection frequency and capability robust. Hence the need to mathematically model the operations of the network for efficient, cheap, and generalizability by introducing an extended mathematical model in future studies to extend the studies in Abdo & Hossain [2]. Technical challenges in port mapping were systematically addressed through distribution and observation, maintaining data integrity across all router instances. Though initial deployments favoured North American and European nodes, subsequent expansions incorporated Asian and South American routers to improve geographic representation. Strict adherence to passive monitoring protocols ensured ethical compliance while enabling robust analysis of routing behaviours. These methodological choices preserved the study’s validity while respecting the inherent constraints of privacy-preserving network research.

# ETHICS STATEMENT

This study followed ethical research guidelines in examining the I2P network. No human subjects, animal testing, or personal data collection were involved. Only network-layer metadata (e.g., node identifiers, tunnel timestamps, country, router performance metrics) from router consoles and directories was analysed. No message content or user activity was accessed. Data was anonymized to ensure privacy. Data collection was designed to minimize network impact, with rates carefully managed to prevent disruptions to normal operations. This approach aligns with ethical standards and best practices for studying anonymous networks.

# CRediT AUTHOR STATEMENT

Siddique Abubakr Muntaka: led the project, contributing significantly to the conceptualization, methodology, experiments, validation, formal analysis, data curation, and writing of the original draft. He also played a crucial role in visualization, supervision, and project administration. Jacques Bou Abdo: served as the research advisor, providing critical reviews and guidance at every step of the project. His contributions were essential in shaping the methodology and ensuring the rigor of the analysis. Kemi Akanbi, Sunkanmi Oluwadare, Faiza Hussein, Oliver Kornyo, and Michael Asante supported the project through various roles, including reviewing and proofreading the manuscript, editing, validation, and offering insights that enriched the final output. Their collective efforts ensured the accuracy and coherence of the study.

# ACKNOWLEDGEMENTS

This research received no specific grant from funding agencies in the public, commercial, or not-for-profit sectors. The foundational work on the computational modeling of the Invisible Internet in Abdo & Hossain [2] was instrumental in this study. Additionally, we would like to thank the University of Cincinnati and the School of Information Technology for providing us with essential resources. The Ohio Cyber Range Institute’s computational resources were crucial in the study experiments. Lastly, Munt IT Solutions cloud hosting resources was very helpful.

# DECLARATION OF COMPETING INTERESTS

The authors confirm that no financial interests or personal relationships could have influenced the research presented in this paper.